\documentclass[showpacs,twocolumn,preprintnumbers,amsmath,amssymb,aps,prl]{revtex4}
\pdfoutput=1
\usepackage{graphicx}
\usepackage{amsmath,amssymb}
\usepackage{epstopdf}
\usepackage{color}%

\usepackage[colorlinks,urlcolor=blue,citecolor=blue]{hyperref}
\def\hhref#1{\href{http://arxiv.org/abs/#1}{arXiv:#1}} 

\begin{document}

\title{Generating Non-perturbative Physics from Perturbation Theory}

\author{Gerald V. Dunne}
\affiliation{Physics Department, University  of Connecticut, Storrs, CT 06269}
\author{Mithat \"Unsal}
\affiliation{Department  of Physics and Astronomy, SFSU, San Francisco, CA 94132}

\begin{abstract}
In a large variety of quantum mechanical systems, we show that the full non-perturbative
expression for energy eigenvalues, containing all orders of perturbative, non-perturbative and quasi-zero-mode terms,  may be generated directly from the perturbative expansion about the perturbative vacuum, combined with a single global boundary condition. 
This provides a dramatic realization of the principle of ``resurgence'', that the fluctuations about different semiclassical saddle points are related to one another in a precise quantitative manner.  
The analysis of quantum mechanics also generalizes to certain calculable regimes of quantum field theory. 

\end{abstract}
\maketitle

It is well known that perturbation theory is generically divergent,
and  that this fact  leads to  
a deep relationship between perturbation theory and non-perturbative physics \cite{Bender:1969si,Lipatov:1976ny,zinnbook}.  
 Here we report a new and deeper level of correspondence. We show that
 perturbation theory contains, in an efficiently coded form, all information about all orders of non-perturbative physics.  The decoding of this information (explained below) requires only that perturbation theory be combined with a global boundary condition that specifies how one degenerate vacuum connects to another. Our result applies to a wide class of  potentials with degenerate vacua, including 
 the double-well (DW) potential  and the periodic Sine-Gordon (SG) potential. These potentials are paradigms for the physics of tunneling, level-splitting and band-splitting, instantons and multi-instantons, and so  have  broad applications in many branches of physics.
  
 Our result extends the previous state-of-the-art, due to Zinn-Justin and Jentschura (ZJJ) \cite{jzj}, who showed that all orders of the perturbative and non-perturbative expansion can be generated using a (conjectured) exact quantization condition, together with two functions, $B(E, g)$ and $A(E, g)$, of the energy $E$ and perturbation parameter $g$. In ZJJ's approach, $B(E, g)$ is equivalent to the perturbative expansion, and $A(E, g)$ is a non-perturbative single-instanton function including fluctuations. We prove our result by showing that  $A(E, g)$ can be deduced immediately from $B(E, g)$; and moreover that ZJJ's quantization condition is the natural global boundary condition in a uniform WKB approach.

For both the DW and SG systems, at leading instanton order the perturbative energy levels are split,  but this is just the tip of the iceberg: there
is also an infinite ladder of higher-order multi-instanton effects. ``Resurgence'' is a formalism that unifies perturbative and non-perturbative analysis, in such a way that the terms in this ladder are intricately related to one another, and can all be expressed in a compact coded form \cite{zinnbook,Argyres:2012vv,Dunne:2012ae}. Here we demonstrate explicitly how resurgence appears in these paradigmatic quantum problems. We show that 
the entire  (non-perturbative) ladder can be generated from perturbation theory about the perturbative vacuum. 

 Although our result is derived in quantum mechanics, it   has implications for asymptotically free quantum field theories  \cite{Argyres:2012vv,Dunne:2012ae,Cherman:2013yfa},  such as two dimensional non-linear sigma models  and four  dimensional  Yang-Mills. 
     A  striking property of  these theories is adiabatic continuity  upon compactification  (with 
       suitable boundary conditions or deformation) \cite{Unsal:2008ch}.  This permits  a continuous connection between the incalculable strong coupling regime and the weak coupling calculable  
          regime, without a phase transitions or rapid cross-over, where the results of our current work also apply.   Such connections between strong- and weak-coupling have broad  physical applications.  Resurgent expansions have also been explored recently for string theory and matrix models \cite{marino,Marino:2012zq,schiappa,Krefl:2013bsa}.

For  potentials with degenerate minima, perturbation theory about any minimum  is
  non-Borel-summable \cite{Brezin:1977gk,Stone:1977au}, and this leads to ambiguous non-perturbative imaginary terms, first seen at the two-instanton level. These  terms are canceled by corresponding ambiguous imaginary terms in 
  the instanton/anti-instanton sector     \cite{Bogomolny:1980ur,ZinnJustin:1981dx}.
   This cancellation, which we refer to as the Bogomolny-Zinn-Justin (BZJ) mechanism,  persists to all non-perturbative orders, producing a real and unambigous result. For example, the  fluctuation about the single-instanton 
  is itself a divergent non-Borel-summable series, generating an ambiguous imaginary term, which is canceled by a term in the three-instanton sector. This infinite ladder of inter-related perturbative and non-perturbative terms is written as a``resurgent trans-series'' expansion of the $N^{\rm th}$ energy eigenvalue \cite{zinnbook}
\begin{eqnarray}
E^{(N)}(g)=\hskip -.1 cm \sum_{\pm}\sum_{k=0}^\infty \sum_{l=1}^{k-1}\sum_{p=0}^\infty c^{\pm}_{k, l, p}\, \frac{e^{-k\frac{S}{g}}}{g^{k(N+\frac{1}{2})}}\hskip -.1 cm  \left(\ln \left[\mp\frac{2}{g}\right]\right)^l \hskip -.1 cm g^{p}
\label{trans}
\end{eqnarray}
In physical terms, the trans-series is a sum over all instanton contributions,   perturbative fluctuations about each instanton sector, and log terms coming from quasi-zero-modes, starting at the two-instanton level. The $\pm$  is correlated with the nature of the quasi-zero mode integration,
yielding $+/-$  for repulsive/attractive   interactions.

We adopt the normalization conventions of ZJJ \cite{jzj}, writing potentials for the DW and SG models as $V(x)=x^2(1-\sqrt{g}\,x)^2/2$, and $V(x)=\sin^2(2\sqrt{g}\, x)/(8g)$. Degenerate vacua are separated by a distance $\sim1/\sqrt{g}$, with a barrier height $\sim 1/g$. When $g=0$ we have isolated harmonic oscillator wells, so for nonzero $g$ it is natural to use a uniform WKB approach \cite{millergood}, writing the wavefunction in terms of a harmonic oscillator wavefunction (the parabolic cylinder function \cite{nist}): $\Psi(x)=D_\nu(u(x)/\sqrt{g})/\sqrt{u^\prime(x)}$. When $g=0$, the ansatz parameter $\nu$ reduces to the integer harmonic oscillator level number $N$.
For example,
\begin{eqnarray}
E(\nu, g)=\sum_{n=0}^\infty  E_n(\nu)\, g^n
\label{eq:pt}
\end{eqnarray}
where the $n^{\rm th}$ perturbative coefficient $E_n(\nu)$ is a polynomial of degree $(n+1)$ in the parameter $\nu$.
When  $\nu$ is an integer, $N$,  this is precisely the standard Rayleigh-Schr\"odinger  perturbative expansion about the $N^{\rm th}$ harmonic level. This expansion is divergent and non-Borel-summable \cite{Brezin:1977gk,Stone:1977au}, and therefore incomplete. This is not surprising, since perturbation theory does not specify a boundary condition to relate one well to another. For the DW, this requires writing the upper (lower) level as an odd (even) function about the barrier midpoint, while for SG we impose a Bloch condition, which can also be specified at a barrier midpoint \cite{nist}. The inherent non-Borel-summability of the perturbative  expansion means that we must analytically continue $g\to g\pm i\epsilon$, which in turn implies that the semiclassical limit requires the asymptotic behavior of the parabolic cylinder functions  slightly off the real axis. This entails a balance between two different exponential terms \cite{uniform}, leading to the  boundary conditions:
 (for DW the $\varepsilon=\pm 1$ refers to the upper/lower level, and for SG, $\theta$ is the Bloch angle):
\begin{eqnarray}
&{\rm DW}&: 
\frac{1}{\Gamma(-\nu)}\left(\frac{e^{\pm i\pi}\, 2}{g}\right)^{-\nu}=-\varepsilon H_0(\nu, g)\, \frac{e^{-S/g}}{\sqrt{\pi g}}
\label{eq:global-dw}\\
&{\rm SG}&: 
\frac{1}{\Gamma(- \nu)}\left(\frac{ 2}{g^2}\right)^{-\nu}  = - \cos \theta \, H_0(\nu, g)\,  \frac{e^{-S/g}}{\sqrt{\pi g}}  \nonumber\\
&& \mp\, i \frac{\pi}{2\, \Gamma(1+ \nu)} \left(\frac{e^{\pm i\pi}\, 2}{g^2}\right)^{+\nu}  \left[H_0 (\nu, g)  \frac{e^{-S/g}}{\sqrt{\pi g}}\right]^2
\label{eq:global-sg}
\end{eqnarray}
where $S=\frac{1}{6}$ for the DW, and $S=\frac{1}{2}$ for SG.
Consider the DW result (\ref{eq:global-dw}). 
The RHS contains the single instanton factor $\xi \equiv e^{-S/g}/\sqrt{\pi g}$, multiplied by  a fluctuation factor $H_0(\nu, g)$. 
When $g=0$, the RHS vanishes, forcing $\nu=N$. For nonzero $g$, 
expanding the Gamma function  shows that $\nu$ is only exponentially close to $N$:
\begin{eqnarray}
\nu_{\rm DW} &=&N
+  \frac{\left(\frac{2}{g}\right)^N}{N!}H_0(N, g)\,\xi 
-  \frac{\left(\frac{2}{g}\right)^{2N}}{(N!)^2}
\left[ H_0 \frac{\partial H_0}{\partial N} \right.
\nonumber \\
&&\hskip -.5 cm \left . +\left(\ln\left(\frac{e^{\pm i\pi}\, 2}{g}\right)-\psi(N+1)\right)H_0^2\right]  \xi^2 + O(\xi^3) 
\label{eq:nu-exp}  \\ 
\cr 
\nu_{\rm SG} &=&N
-\cos\theta \,  \frac{\left(\frac{2}{g}\right)^N}{N!}H_0(N, g)\,\xi 
-  \frac{\left(\frac{2}{g}\right)^{2N}}{(N!)^2}
\left\{\left[ H_0 \frac{\partial H_0}{\partial N} \right. \right.
\nonumber \\
&&\hskip -1.5 cm \left .\left.  +\left(\ln\left(\frac{2}{g}\right)-\psi(N+1)\right)H_0^2\right] \cos 2\theta \pm i\pi H_0^2\right\} \xi^2 + O(\xi^3) \qquad 
\label{eq:nu-sg}
\end{eqnarray}
Notice the appearance of the $\ln g$ factors at the two-instanton level: $O(\xi^2)$. Subsituting the expansion (\ref{eq:nu-exp}, \ref{eq:nu-sg})  into the energy (\ref{eq:pt}) generates the full trans-series (\ref{trans}). 

The conditions  (\ref{eq:global-dw}, \ref{eq:global-sg}) are identical to the conjectured exact quantization conditions of ZJJ, but with a  different interpretation. 
(To convert to the notation of ZJJ \cite{jzj}, write $B\equiv \nu+\frac{1}{2}$, and $e^{-A/2}\equiv \sqrt{\pi g}\, \xi\, H_0$.)
 ZJJ view $\nu=\nu(E, g)$ as a function of $E$, rather than $E$ as a function of $\nu$, as in (\ref{eq:pt}); and they also express $H_0=H_0(E, g)$ as a function of energy. The ZJJ strategy is to separately compute the perturbative function $B(E, g)$, and the non-perturbative function $A(E, g)$, and then insert them into the exact quantization condition (\ref{eq:global-dw}, \ref{eq:global-sg}), obtaining an implicit transcendental expression for $E$ as a function of $g$, whose small $g$ expansion yields the trans-series (\ref{trans}). An advantage of our uniform WKB approach is that it reveals a simple relation between the functions $B$ and $A$.

To illustrate this, we first note that the perturbative expansion $E=E(\nu, g)$ in (\ref{eq:pt})  agrees precisely,  using  the identification $B\equiv \nu+\frac{1}{2}$,  with the inversion of the ZJJ expressions for $B=B(E, g)$ in \cite{jzj}:
\begin{eqnarray}
E_{\rm DW}(B, g)&=&B-g\left(3B^2+\frac{1}{4}\right)-g^2\left(17 B^3+\frac{19}{4}B\right)
\nonumber\\
&&\hskip -1cm -g^3\left(\frac{375}{2}B^4+\frac{459}{4}B^2+\frac{131}{32}\right)-\dots
\label{eq:bdw2}
\\
E_{\rm SG}(B, g)&=&
B-g\left(B^2+\frac{1}{4}\right)-g^2\left(B^3+\frac{3 B}{4}\right) \nonumber\\
&&
-g^3\left(\frac{5
   B^4}{2}+\frac{17 B^2}{4}+\frac{9}{32}\right)-\dots
\label{bsg2}
\end{eqnarray}
Next, we use these expressions for $E(B, g)$ to write ZJJ's non-perturbative function $A(E, g)$ \cite{jzj} as $A(B, g)$:
\begin{eqnarray}
A_{\rm DW}(B, g)&\hskip -.2 cm =&\hskip -.2cm \frac{1}{3g}+g\left(17B^2+\frac{19}{12}\right)+g^2\left(125 B^3+\frac{153 B}{4}\right)
\nonumber\\
&& \hskip -1cm +g^3\left(\frac{17815}{12}B^4+\frac{23405}{24}B^2+\frac{22709}{576}\right)+\dots
\label{eq:adw2}
\\
A_{\rm SG}(B, g)&=&\frac{1}{g}+g\left(3 B^2+\frac{3}{4}\right) +g^2
\left(5 B^3+\frac{17 B}{4}\right)
\nonumber\\
&& + g^3 \left(\frac{55 B^4}{4}+\frac{205 B^2}{8}+\frac{135}{64}\right) +\dots 
 \label{asg2}
\end{eqnarray}
Inspection of  (\ref{eq:bdw2}, \ref{bsg2}, \ref{eq:adw2}, \ref{asg2}) reveals the simple relations:
\begin{eqnarray}
\frac{\partial E_{\rm DW}}{\partial B}&=&-6\, B\, g-3g^2\frac{\partial A_{\rm DW}}{\partial g}
\label{eq:dw-magic}\\
\frac{\partial E_{\rm SG}}{\partial B}&=&- 2B\, g-g^2\frac{\partial A_{\rm SG}}{\partial g}
\label{eq:sg-magic}
\end{eqnarray}
Similar relations arise by inverting and re-expanding  the expressions for $B(E, g)$ and $A(E, g)$ in \cite{jzj} for the Fokker-Planck and $O(d)$ AHO potentials.
We can write  (\ref{eq:dw-magic}, \ref{eq:sg-magic}) as the general expression:
\begin{eqnarray}
\frac{\partial E}{\partial B}=- \frac{g}{2S}\left(2B+g\frac{\partial A}{\partial g}\right)
\label{eq:general-magic}
\end{eqnarray}
where $S$ is the numerical coefficient of the instanton action in  $\xi\equiv e^{-S/g}/\sqrt{\pi g}$.  Equation (\ref{eq:general-magic})  is our main result.
This relation (\ref{eq:general-magic}) has a dramatic computational consequence:  we can deduce 
$A=A(B, g)$ immediately from the perturbative expansion (\ref{eq:pt}) for $E(B, g)$: thus, the instanton computation for $A(B, g)$ is actually unnecessary, as it is already encoded in the perturbative expression for $E(B, g)$. 
Furthermore, the overall normalization factor $S$ is encoded in the leading non-alternating  large-order  growth $n!/(2S)^n$ of the ground state perturbative expansion, or is easily computed from the single instanton integral. 
This proves our claim that the entire trans-series can be generated from perturbation theory, together with the global boundary condition. The relation (\ref{eq:general-magic})   is not at all obvious in the ZJJ expressions where $B$ and $A$ are written as functions of $E$, because it requires an inversion and re-expansion. For the DW, a related form of (\ref{eq:dw-magic}) was noted in \cite{alvarez}, but its physical meaning and consequences were not pursued.

We now explain the resurgent origin, and implications,  of this general relation (\ref{eq:general-magic}). 
The key step is to note that the trans-series (\ref{trans}) arises from combining the formal perturbative expansion (\ref{eq:pt}) with the global condition that determines $\nu$ to be exponentially close to its perturbative value $N$, as {\it e.g.} in (\ref{eq:nu-exp}, \ref{eq:nu-sg}). Therefore, we expand
\begin{eqnarray}
E_N(g)=E(N, g)+(\delta \nu) \left[\frac{\partial E}{\partial \nu}\right]_{N}
+\frac{(\delta \nu)^2}{2} \left[\frac{\partial^2 E}{\partial \nu^2}\right]_{N}+\dots
\label{eq:expansion}
\end{eqnarray}
Each correction term involves two factors. One is a perturbative factor of derivatives w.r.t. $\nu$ of  the perturbative expression $E(\nu, g)$, evaluated at $\nu=N$. The other is a non-perturbative factor of powers of the non-perturbative shift $\delta \nu$ of $\nu$ from its integer value $N$, as in (\ref{eq:nu-exp}, \ref{eq:nu-sg}).
But $\delta \nu$ is expressed entirely in terms of $\xi$ and $H_0$, and because of the general relation (\ref{eq:general-magic}), each of these is easily deduced from the perturbative expansion. From (\ref{eq:general-magic}) and the identification $e^{-A/2}\equiv \sqrt{\pi g}\, \xi\, H_0$, the instanton fluctuation factor is derived from $E(B, g)$ as:
\begin{eqnarray}
H_0=\exp\left[S\int_0^g \frac{dg}{g^2}\left(\frac{\partial E}{\partial B}-1+\frac{B\, g}{S}\right)\right]
\label{eq:h0}
\end{eqnarray}
For example,
\begin{eqnarray}
H_0^{\rm DW}(N, g)&=& 1-\frac{\left(102 N^2+102 N+35\right)}{12} g 
\label{eq:dwh0}\\
&&\hskip -3cm +\frac{\left(10404 N^4+2808 N^3-9456 N^2-11868
  N-3779\right)}{288} g^2 +\dots \nonumber \\
 H_0^{\rm SG}(N, g)&=& 
 1-\frac{3\left(2 N^2+2 N+1\right)}{4} g \nonumber\\
 &&\hskip -2.5cm +
  \frac{ \left(36 N^4-8 N^3-48 N^2-92
   N-35\right)}{32} g^2+\dots
   \label{eq:sgh0}
   \end{eqnarray}
As an immediate application, consider the first order in the instanton expansion, which gives the level/band splitting. For the $N^{\rm th}$ level, we have
\begin{eqnarray}
\Delta E_N(g)\sim 2 \frac{\left(\frac{2}{g}\right)^N}{N!}\, \frac{e^{-S/g}}{\sqrt{\pi g}} \, 
F_0(N, g)
\label{eq:one-fluc}
\end{eqnarray}
The leading factor is the familiar text-book single-instanton result, while the non-trivial fluctuation factor, $F_0(N, g)=H_0(N, g) \left[\frac{\partial E}{\partial \nu}\right]_{N}$, can be deduced entirely from the perturbative expansion $E(\nu, B)$, using (\ref{eq:pt}, \ref{eq:h0}):
\begin{eqnarray}
F_0^{\rm DW}(N, g)&=&
1-\frac{(102 N^2+174 N+71)}{12} g +O(g^2) \qquad 
\label{eq:dw-inst-fluc}
\\
F_0^{\rm SG}(N, g)&=&1-\frac{\left(6 N^2+14 N+7\right)}{4}
   g +O(g^2)
\label{eq:sg-inst-fluc}
   \end{eqnarray}
This DW result agrees with available ($N=0$)  results from instanton calculus
 \cite{ZinnJustin:1979db,Wohler:1994pg} which is obtained by evaluating the Feynman diagrams  up to the two-loops order in the instanton background. Similarly,  for SG we agree with results from the Mathieu equation for any $N$ in \cite{nist,dingle}. It is striking that  our formalism correctly produce the perturbative fluctuations around an instanton  background {\it without} evaluating the Feynman diagrams in that background! 
This is the power of resurgence.

Now consider the two-instanton sector. From  (\ref{eq:nu-exp}, \ref{eq:nu-sg}),  the  first imaginary term \footnote{Note that the imaginary term is independent of the parity factor $\varepsilon$ for the DW, and of the Bloch angle $\theta$ for SG; this is crucial because these imaginary terms must cancel against terms coming from Borel summation of perturbation theory, and perturbation theory is independent of these parity and Bloch angle parameters.} enters at $O(\xi^2)$ from  $\delta \nu$:
\begin{eqnarray}
{\rm Im}\, E_N(g) \sim \frac{2^{2N}}{(N!)^2} \frac{e^{-2S/g}}{g^{2N+1}} H_0^2(N, g) \left[\frac{\partial E}{\partial \nu}\right]_{N}
\label{eq:imag}
\end{eqnarray}
A dispersion relation for the energy gives the leading large order growth of the perturbative coefficients, as a function of the level number $N$:
\begin{eqnarray}
c_n^{(N)} &\sim &\frac{1}{\pi}\int_0^\infty \frac{dg}{g^{n+1}}\, {\rm Im}\, E_N(g)
\label{eq:large-order-N}\\
&\sim &
-\frac{1}{\pi} \frac{2^{2N}}{(N!)^2} \frac{\Gamma(n+2N+1)}{(2S)^{n+2N+1}}
\left[1+O\left(\frac{2S}{n+2N}\right)\right]
\nonumber
\end{eqnarray}
\begin{figure}[htb]
\includegraphics[scale=.44]{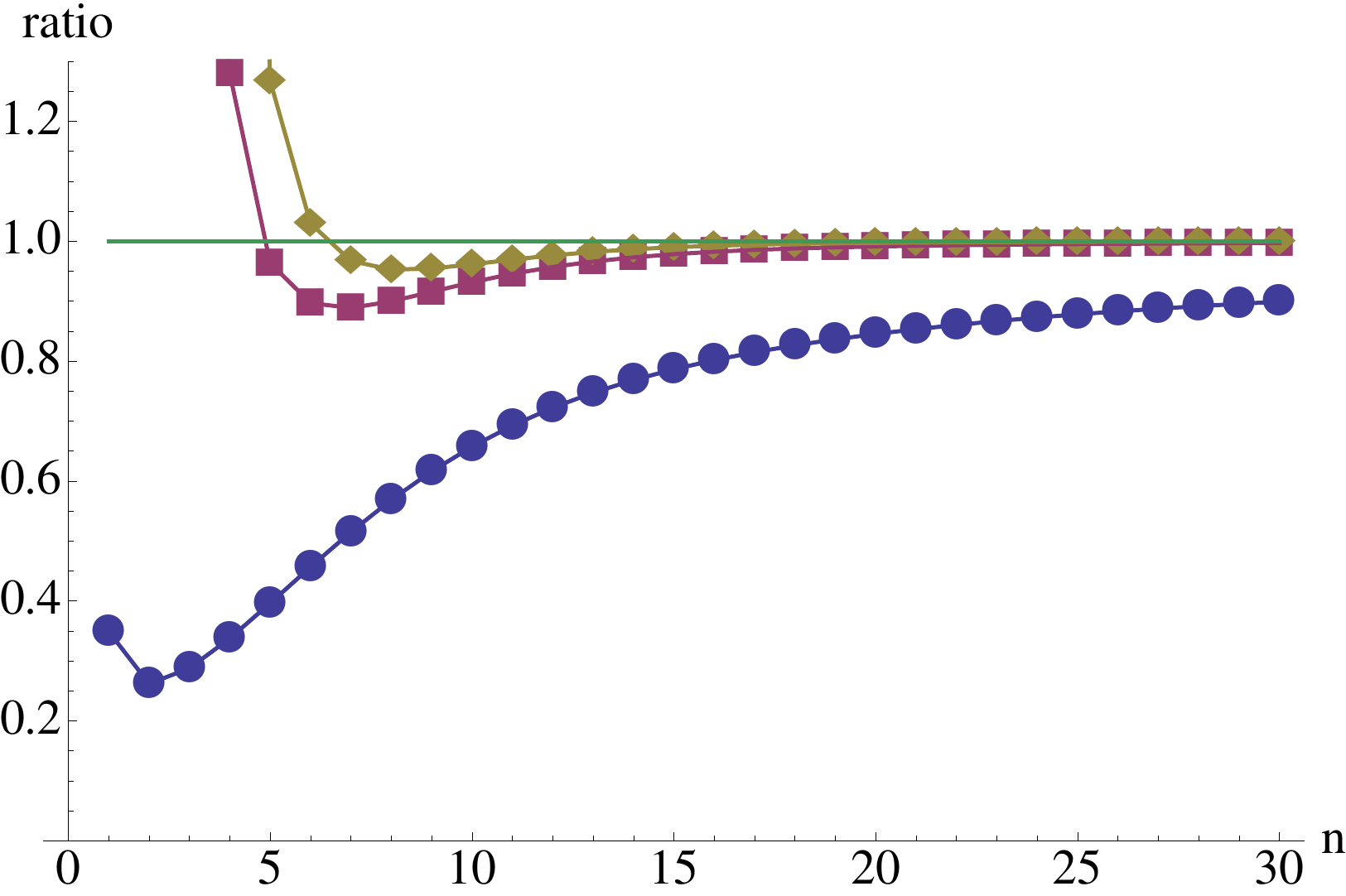}
\includegraphics[scale=.44]{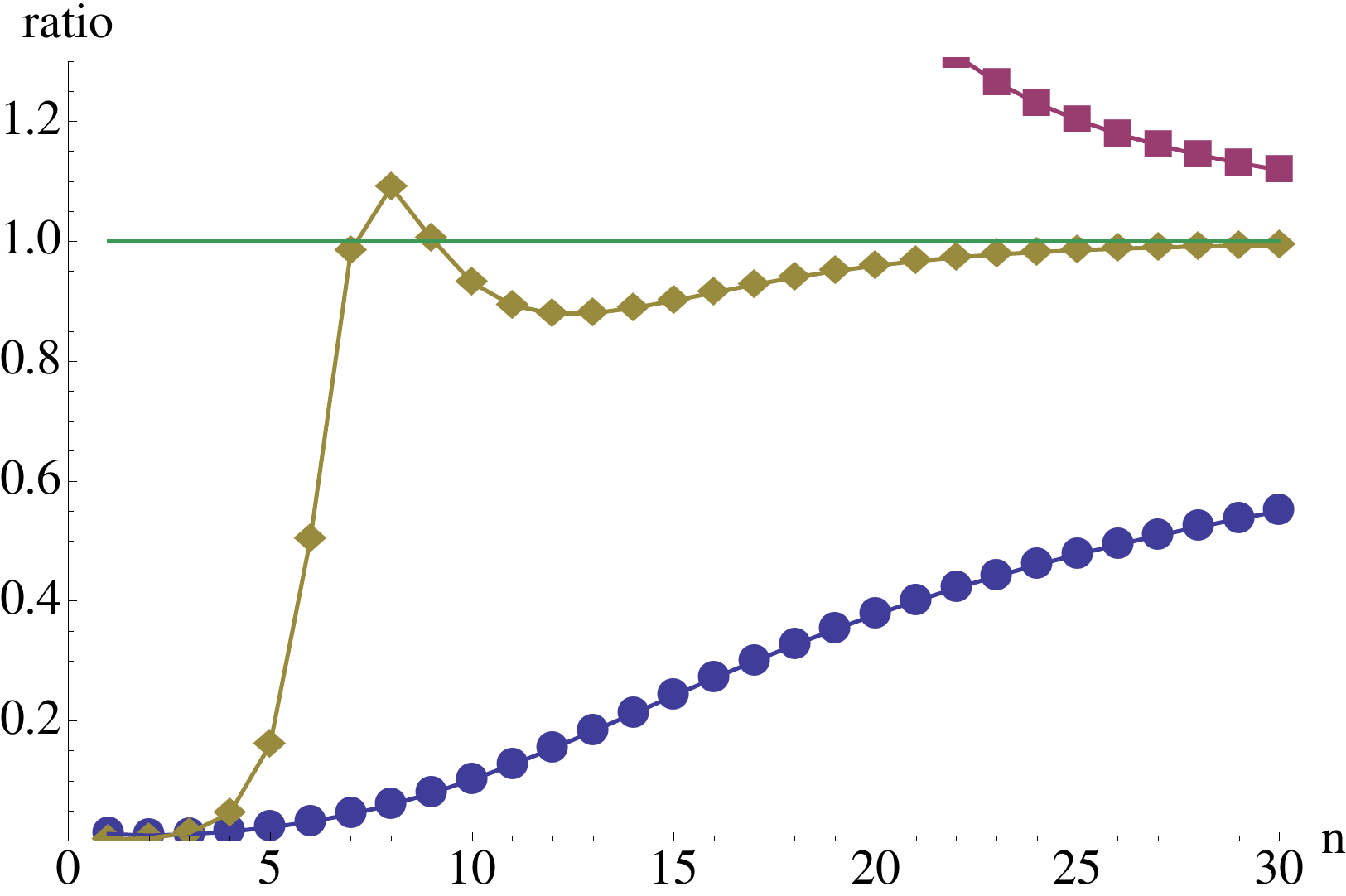}
\caption{Plots of the ratio of the exact perturbative coefficients to the large-order growth (\ref{eq:large-order-N}), for the double-well potential, for the $N=0$  ground state (upper plot) and $N=1$ first excited state (lower plot). The solid (blue) circles, (red) squares and (gold) diamonds denote the leading, sub-leading and sub-sub-leading large $n$ behavior, respectively, including increasing information from (\ref{eq:h0}, \ref{eq:imag}, \ref{eq:large-order-N}) concerning the fluctuations about the instanton/anti-instanton sector. }
\label{fig1}
\end{figure}
Fig. \ref{fig1} 
shows excellent  agreement with the numerically computed large order growth,  from the first 30 terms in the perturbative expansion, also including sub-leading corrections to (\ref{eq:large-order-N}) that come from the fluctuations about the instanton/anti-instanton sector, for  the DW  potential, using (\ref{eq:dwh0}, \ref{eq:sgh0}, \ref{eq:imag}, \ref{eq:large-order-N}). Analogous results apply to the SG potential.

This large-order growth is not directly evident from the perturbative expansion (\ref{eq:pt}), as it involves the large order growth of the polynomials $E_n(B)$, for different values of $B$. The result (\ref{eq:large-order-N}) requires the input of the general relation (\ref{eq:general-magic}), and its immediate corollary (\ref{eq:h0}). 
An interesting example is  the Fokker-Planck potential \cite{jzj},  the SUSY QM double-well potential,
with ground state energy that is perturbatively  zero. Here the global condition leads to $B\sim N+\delta B$, and  SUSY is broken non-perturbatively at the 2-instanton level: $E^{(0)}\sim \frac{1}{2\pi} e^{-\frac{1}{3g}}$. The ground state perturbation expansion is convergent [equal to 0], and correspondingly the 2-instanton term has no imaginary part  for $N=0$. For all excited states $N\geq 1$, the perturbative expansion is non-Borel-summable, with associated imaginary parts in the 2-instanton sector. All this information is encoded in the perturbative expansion (\ref{eq:pt}), via  relation (\ref{eq:general-magic}).

A deeper manifestation of the relation (\ref{eq:general-magic}) comes from consideration of  the high orders of the perturbative fluctuations $F_0(N, g)$ about the single-instanton term in (\ref{eq:one-fluc}). Like the perturbation around the vacuum, these fluctuations are divergent and non-Borel-summable, producing ambiguous non-perturbative imaginary terms that must be cancelled by other terms in the trans-series. The leading large order growth of $\frac{\partial E}{\partial \nu}$ can be deduced from (\ref{eq:large-order-N}):
\begin{eqnarray}
\frac{\partial c_n^{(N)}}{\partial N} 
\sim 
-\frac{1}{\pi} 
\frac{2^{2N+1}}{(N!)^2} \frac{\Gamma(n+2N+1)}{(2S)^{n+2N+1}} \ln(n+2N+1)
\label{eq:deriv}
\end{eqnarray}
Notice the $(n! \ln n)$ large order growth, different from the conventional $n!$  behavior.
The origin of this lies in the resurgent structure of the trans-series. These perturbative fluctuations about the single instanton produce an imaginary term $\sim e^{-2S/g}$, which when combined with the single instanton factor $\xi$, become $O(e^{-3S/g})$, which must be cancelled by a term in the three instanton sector, coming from instanton/instanton/anti-instanton quasi-zero modes terms, as encoded in the ``resurgence triangle'' \cite{uniform,Dunne:2012ae}. Indeed, in the three-instanton sector  of the trans-series (\ref{trans}), because of the $ ( \ln (-2/g))^2$ term, the trans-series has an imaginary part proportional to $e^{-3S/g} \ln g$, and this corresponds to this novel  $(n! \,\ln n)$ large order growth in the perturbative fluctuations around the single-instanton. Again, this correspondence relies crucially on the general relation (\ref{eq:general-magic}). 

To conclude, we have shown that the remarkable results of Zinn-Justin and Jentschura \cite{jzj} can be extended even further:  the full non-perturbative trans-series expansion of energy eigenvalues is encoded already in the perturbative expansion  $E(\nu, g)$.
Practically speaking, this eliminates the necessity of computing the complicated multi-instanton function $A(E, g)$ in the ZJJ approach, as it can be deduced immediately from  (\ref{eq:general-magic}). We have further shown that this relation is itself a statement of resurgence, required in order to produce the correct cancellations between the zero and two-instanton sectors, and the one- and three-instanton sectors, and so on. 
Ultimately, our result is a manifestation of resurgent structure \cite{Berry,marino,Marino:2012zq,schiappa,Dunne:2012ae,bdu} in the quantum path integral: the fluctuations about different saddle points (vacuum and different multi-instanton sectors) are quantitatively related so tightly that the entire trans-series, to all non-perturbative orders, can be encoded in terms of just the perturbative saddle point. 

By the idea of adiabatic continuity mentioned in the Introduction,  resurgent  structure  is also present in the weak coupling calculable regimes of quantum field theory.  It is  important to understand  how this result can be generalized to strongly coupled quantum fields.

We acknowledge support from DOE grants  DE-FG02-92ER40716 and  DE-FG02-13ER41989 (GD), and  DE-FG02-12ER41806 (M\"U).

\end{document}